\documentclass[reprint]{revtex4-2}

\usepackage{dcolumn}
\usepackage{bm}
\usepackage{indentfirst}
\usepackage{mathrsfs}
\usepackage{comment}
\usepackage{amsfonts}
\usepackage{amssymb}
\usepackage{dsfont}
\usepackage{euscript}
\usepackage{graphicx}
\usepackage{wrapfig}
\usepackage{mathtools}
\usepackage{pifont}
\usepackage[dvipsnames]{xcolor}
\usepackage{tensor}
\usepackage{amsmath,latexsym,esint}
\usepackage{cmap}
\usepackage[figurename=Figure,font=footnotesize]{caption}
\usepackage{subcaption}
\usepackage{hyperref}
\usepackage{float}
\usepackage{xcolor}
\usepackage[toc,page]{appendix}

\begin{document}
	\preprint{}
	\title{
    Gravielectric and gravimagnetic fluxes in nutty black holes}
    
	\author{Dmitri Gal'tsov}
	\email{galtsov@phys.msu.ru}
	\author{Rostom Karsanov}
	\email{karsanovrz@my.msu.ru}
	\affiliation{Faculty of Physics, Moscow State University, 119899, Moscow, Russia
	}

\begin{abstract}
 We introduce the gravielectric (GE) and gravimagnetic  (GM) fields  in stationary spacetime using the Komar two-form and its dual. This opens the way to extend the Komar-Tomimatsu  derivation of mass formulas to a more detailed picture in terms of the local lines of force.  We show that Misner strings (MS) carry singular GE and GM fluxes  connecting  the horizon  and the asymptotic zone. Moreover, MS are laterally transparent, so field lines can flow in and out of the bulk. This explains why the usual Komar mass integrals around the Misner strings in the Taub-NUT vacuum solution are negative: the pattern of field lines shows that they flow onto the string from the horizon, so it is necessary to calculate the incoming (positive) but not the outgoing Komar fluxes. This incoming flux is then turned back to the horizon through the Misner strings, realizing the closed circuit without sources. So Misner strings are massless empty tubes, but not rigid rods of negative mass. Similarly, GM field lines can connect positively and negatively charged regions of the horizon, generating, for example, the gravimagnetic dipole moment of the Kerr metric.  
\end{abstract}
	
\maketitle

\section{Introduction} 

    Electromagnetic duality \cite{Misner:1957mt} is a fundamental concept in theoretical physics, representing the symmetry in which electric and magnetic fields can be interchanged, along with charges and monopoles, without changing Maxwell's equations. It readily generalizes to curved space electrodynamics and Yang-Mills theory \cite{Deser:1976iy} 
    and is crucial for understanding dualities in quantum field theories and string theory. 
This duality implies that in many scenarios, a strongly coupled electric theory can be mapped to a weakly coupled magnetic theory (S-duality), offering powerful tools in calculations. It extends to higher linear spins \cite{Bunster:2006rt}, most importantly, to linearized gravity \cite{Szekeres:1971ss,Maartens:1997fg,Nieto:1999pn,Mashhoon:2003ax,Henneaux:2004jw, Barnich:2008ts,Ramos:2010zza,Bunster:2012km,Boos:2021suz},   but the extension to full non-linear gravity was questioned \cite{Ellwanger:2001uq,Deser:2005sz,Monteiro:2023dev}. Recently, it was shown that using the frame definition of duality transformation one can express the self-dual part of the curvature tensor \cite{Kol:2022bsd} in the form bearing the elements of exact S-duality, but this does not solve the full problem.  Gravitational S-duality was also extensively discussed in supergravity \cite{Argurio:2008zt,Argurio:2009xr,Moutsopoulos:2009ia,Dehouck:2011xt} and string theory \cite{Hull:2000zn,Hull:2001iu,Hull:2023iny,Kol:2023yxd}. 

The cornerstone of non-linear gravitational S-duality is the Taub-NUT solution of the vacuum Einstein equations \cite{Newman:1963yy,Misner:1963fr,Miller:1971em,Sackfield71,Dowker:1974znr,Mcguire:1975wq,Ramaswamy:1981,Mueller:1985ij,Bini:2003bm}. This solution, meanwhile, has severe problems associated with Misner strings. Dropping the Misner time identification which eliminates Misner strings at the price of introducing  closed timelike curve everywhere, one can adopt Bonnor's view \cite{Bonnor:1969ala}, that Misner strings are true line singularities. Developing this idea, it was shown \cite{Clement:2015cxa,Clement:2016mll} that one of the long-standing problems --- closed timelike curves in the neighborhood of MS --- can be partially solved, since these lines are not geodesics, while MS itself are geodesically transparent. Mass formulas and the thermodynamics of nutty black holes were developed \cite{Clement:2017otx,Clement:2019ghi, Hennigar:2019ive,Wu:2019pzr,Chen:2019uhp,Awad:2022jgn}. Meanwhile, Bonnor's interpretation was criticized \cite{Manko:2005nm} arguing that singular MS have unphysical matter sources.
Based on linearized theory, Bonnor proposed that Misner strings are massless sources of rotation. However, in nonlinear theory, the Komar mass integral over the sidewall of Misner strings is negative, which was interpreted in \cite{Manko:2005nm} as an unphysical negative mass of the string. Here, we show that this interpretation can be challenged. Our argument is based on the approach of localizing Komar integrals using the concept of Komar field lines.
 
In linearized general relativity the curvature tensor can be split into electric and magnetic sectors leading to Maxwell-like equations for GE and GM fields. Here we suggest the definition applicable in the non-linear regime for stationary spacetime, which is directly related to the Komar two-form associated with stationarity.  Using them we can extend the Komar-Tomimatsu integral approach   to more informative picture of GE and GM lines of force.
The present paper generalizes to gravity our recent results \cite{Galtsov:2026ndi} on  the short-range electromagnetic structures of the charged nutty black holes.

\section{The setup} 

We start with the Taub-NUT metric which is an exact solution of the vacuum Einstein equations with two parameters: the mass $m$,  and the NUT parameter $n$: 
 \begin{align}\label{TaubNUT}
    	&ds^2=-\frac{\Delta}{\Sigma}(dt+2n\cos \theta d\varphi)^2+\frac{\Sigma}{\Delta}dr^2+\Sigma d\Omega^2,\\
    	&\Delta=r^2-2mr-n^2,\qquad \Sigma=r^2+n^2.
    \end{align} 
This metric has the timelike Killing vector $k^\mu=\delta^\mu_t$, which becomes null on the horizon $\Delta(r_H)=0$:
\begin{equation}
    r_H=m+\sqrt{m^2+n^2}.
\end{equation}

In contrast to the Schwarzschild case,
the metric has no singularity at $r=0$. For definiteness, we assume $m>0,\,n>0$. 

This metric also has a rotational Killing vector $m^\mu=\delta^\mu_\varphi$
and admits a canonical representation in Weyl-Papapetrou coordinates
\begin{equation}
  ds^2 = -F(dt-\omega d\varphi)^2 + F^{-1}[e^{2k}(d\rho^2+dz^2)+\rho^2d\varphi^2],
\end{equation}
related as $\rho=\sqrt{\Delta}\sin\theta,\;z=(r-m)\cos\theta$. The polar axis $\rho=0$ is divided into three rods \cite{Harmark:2004rm,Clement:2019ghi}: the horizon rod $|z|<r_H$ and two Misner string rods $z>r_H,\; z<-r_H$ (north and south strings). 

Each rod is a Killing horizon of a certain linear combination of $ k^\mu$ and $m^\mu$ and is characterized by a constant two-dimensional directional vector  whose components are related to surface gravity $\kappa$, and angular velocity $\Omega$ of the rod.   For the Taub-NUT metric the horizon is non-rotating, $\Omega_H=0$, while Misner strings are rotating with velocities $\Omega_\pm=\mp 1/2n$.

Although Misner strings appear as one-dimensional defects, they are actually two-dimensional cylinders with finite lateral surface area per unit length \cite{Clement:2019ghi}:
\begin{equation}\label{A}
  {\cal{A}} =\oint d\varphi  \sqrt{|g_{zz}g_{\varphi\varphi}|} =2\pi  \vert e^k\omega \vert.  
\end{equation}
Thus, the image of infinitely thin hollow tube  is a more accurate geometric representation of Misner strings than a line. 

\section{GE and GM fluxes in Misner strings}
Let us show that the MS cylinders carry fluxes of gravielectric and gravimagnetic fields defined as follows.
Consider the Komar formula for the mass of the stationary gravitational field in the volume $V$ surrounded by the closed two-surface $\partial V$  
given by the integral
\begin{equation}\label{Mkomar}
  M=-\frac1{8\pi} \oint_{\partial V } \star dk,
\end{equation}
where $k=g_{t\nu}dx^\nu$ is the Komar one-form corresponding to the timelike Killing vector
$k^\mu=\delta^\mu_t$, and the star denotes the Hodge dual.  For the Levi-Civita symbol we use a  convention $\epsilon_{tr\theta \varphi}=-\epsilon^{tr\theta \varphi}=1$. 

The Komar integral for the dual gravimagnetic charge is\begin{equation}\label{Nkomar}
  N=\frac1{8\pi} \oint_{\partial V }  dk.
\end{equation} 

Commuting the covariant derivatives and using Killing equations one can derive Maxwell equations  and Bianchi identities for Riemann tensor in terms of Komar two-form and its dual \cite{Bossard:2008sw,Kol:2020zth}:
\begin{equation}\label{GR EOM K}
    \begin{aligned}
        d&\star K=2 \star dx^\mu  R_{\mu\nu}k^\nu ,\\
    &dK=-\frac{1}{2}k^\sigma R_{[\rho\mu\nu]\sigma}dx^\rho\wedge dx^\mu\wedge dx^\nu,
    \end{aligned}
\end{equation}
which for the vacuum solutions and a torsion-free theory reduce  to free equations
\begin{equation*}
	d\star K=0, \qquad dK=0,
\end{equation*}
 analogous to the sourceless Maxwell equations and Bianchi identities, written in terms of the Maxwell two-form and its dual:
\begin{equation}
	d\star F=0,\qquad dF=0.
\end{equation}
But this is true only in the regular part of spacetime, from which the Misner string singularities must be cut out, surrounding them, e.g., by thin cylindrical surfaces. To discover singular fluxes we have to consider the potentials for $K$ and $\star K$. For the first it is simply the Komar one-form
\begin{equation}\label{1-form g}
    	k=-\frac{\Delta}{\Sigma}(dt+2n\cos \theta d\varphi),
    \end{equation}
    and it leads to the bulk Komar two-form
    \begin{equation}\label{Komar 2 form}
    	K=2m\frac{\tilde{\Delta}}{\Sigma^2}(dt+2n\cos \theta  d\varphi)\wedge dr+2n\frac{\Delta}{\Sigma}\sin \theta d\theta \wedge d\varphi,
    \end{equation}
    where we have introduced 
    \begin{equation}
    	\tilde{\Delta}=r^2+2\frac{n^2}{m}r-n^2.
    \end{equation} 
    In view of Misner string singularity, we still have to look for the singular contribution to the Komar two-form like in the electromagnetic case \cite{Galtsov:2026ndi}.  Using the fact that $d\varphi  $ is not closed on the polar axis \cite{Galtsov:2026ndi}
\begin{equation}\label{ddphi}
dd\varphi=2\pi\,\delta^2(\mathbf{x})\,dx\wedge dy,
	\end{equation}
    where $x,y$ are the local Cartesian coordinates in the plane orthogonal to $z$ axis, one finds that the exterior derivative $dk$
has in addition the singular part
  \begin{equation}
      dk=K+{\cal K},
  \end{equation} proportional to the NUT charge:
    \begin{align} \label{NF} 	\mathcal{K}^\pm= \mp 4\pi n\frac{\Delta}{\Sigma} \delta(x)\delta(y)dx \wedge dy. 
    \end{align} 

    The dual quantity $\star K
    $ whose reverse flux determines the Komar mass in accordance with \eqref{Mkomar} is given by
\begin{equation}\label{dual Komar 2 form}
 \star K=2n\frac{\Delta}{\Sigma^2}(dt+2n\cos \theta  d\varphi)\wedge dr-2m\frac{\tilde{\Delta}}{\Sigma}\sin \theta d\theta \wedge d\varphi.
    \end{equation}
    
The corresponding potential one-form $\tilde{k}$ such that $\star K=d\tilde{k}$ is equal to
\begin{equation}\label{1-form dual g}
    	\tilde{k}=\frac{m}{n}\frac{\tilde{\Delta}}{\Sigma}\left(dt+2n\cos \theta d\varphi\right).
    \end{equation} 
  Exactly as above, one finds  that $d\tilde{k}$ apart of the bulk field  
\eqref{dual Komar 2 form} contains a singular term
\begin{equation}
    d\tilde{k}=\star K+\tilde{{\cal K}},
\end{equation} 
where the second term is localized on the Misner strings
 \begin{align}\label{MF}
\tilde{{\cal K}}^\pm= \pm 4\pi m\frac{\tilde{\Delta}}{\Sigma}\delta(x)\delta(y)dx \wedge dy.
    \end{align} 
    
These singular fluxes have the following features. GE fluxes are constant at infinities $z\rightarrow\pm \infty$ and are incoming there, as we will see shortly. They increase approaching the horizon. GM fluxes are also constant at infinity, but vanish at the horizon

In contrast with the  electromagnetic case \cite{Galtsov:2026ndi} where the corresponding electric and magnetic fluxes are self-dual, here S-duality holds only as $r\to \infty$.

\section{Balance of fluxes}

So far it was assumed that to get regular spacetime it is enough to cut the Misner string from the lateral side by cylindrical surfaces. Then calculating the flux of $\star K$ through the lateral surface led to a negative mass of the MS strings. But presence of singular fluxes of GE and GM fields inside the strings and finiteness of their transverse sections demands to include these sections too in the definition of the closed surface surrounding the strings. The total boundaries $\partial V_{\pm}$  consist of the sum of the surfaces orthogonal to the string at infinity $S_\perp^\infty$, at the horizon $S_\perp^H $ and the lateral surface $S_L$: 
\begin{equation}\partial V_{MS}=S_\perp^\infty \cup S_\perp^H \cup S_L. 
     \end{equation}

Motivated by the form of the Komar integrals \eqref{Mkomar} and \eqref{Nkomar} we define the fluxes of the GE and GM fields in the bulk as
\begin{equation}\label{bulk fluxes}
\Phi=-\oint_{S} \star K,\qquad\tilde{\Phi}=\oint_{S}  K.
\end{equation}
 Here $S$ is either a lateral surface of cylinders $S_L$ either sphere of some radius $S_r$,
and we will assume that the normals to the sphere and cylinders around the strings are all directed outward of these surfaces.  For the fluxes through the cross sections of the strings then we have analogously a definition
\begin{equation}\label{string fluxes}
	\Phi_\perp=-\oint_{S } \tilde{\mathcal{K}},\qquad\tilde{\Phi}_\perp=\oint_{S }  {\cal K},
\end{equation}
and the normal vectors being directed outside from the horizon to infinity along the strings.

Total flux through the surface $S_\perp$ orthogonally intersecting the strings with the normal in outer direction (i.e. towards increasing $|z|$) at the point $r$ on the strings will read
\begin{equation} \Phi_\perp^\pm(r)=-4\pi m \frac{\tilde{\Delta}(r)}{\Sigma(r)}
\end{equation}

The fluxes through the transverse sections will then be given by $\Phi_\perp^\pm(\infty)$ and $\Phi_\perp^\pm(r_H)$ respectively, while the flux over the lateral surface $\Phi_L$ must be computed using the expression for the bulk integral \eqref{bulk fluxes}.
Then the results of calculation will read
\begin{align}
	&\Phi_\perp^\pm(\infty) =-4\pi m,\\
&\Phi_\perp^\pm(r_H)=-4\pi\sqrt{m^2+n^2},\\
	&\Phi_L^\pm =-4\pi\frac{n^2}{r_H}=4\pi(m-\sqrt{m^2+n^2}).
	\end{align}
The masses of Misner strings according to the Komar definition \eqref{Mkomar} will read
\begin{equation}
    M_\pm= \frac{1}{8 \pi}[\Phi_\perp^\pm(\infty)-\Phi_\perp^\pm(r_H)+\Phi_L^\pm] =0.
\end{equation}
 So the Misner strings are massless indeed, as was originally assumed by Bonnor! 

\begin{figure}[H]
 	\centering
\includegraphics[width=0.79\linewidth]{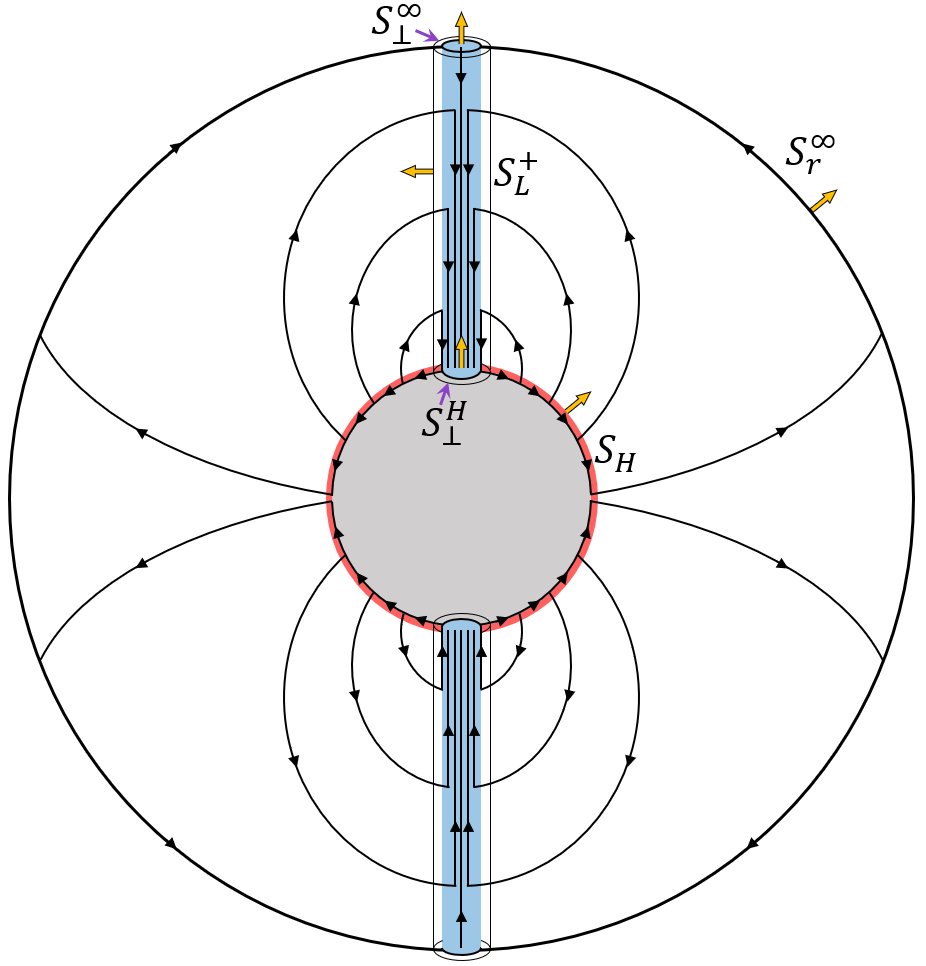}
\vspace{-0.3 cm}
 	\caption{Schematic picture of GE lines of force in Taub-NUT solution. Yellow arrows indicate the directions of the corresponding normal vectors to the surfaces of integration. The sphere $S_r$ is supposed to have large (infinite) radius where it intersects the Misner string at the circles $S_\perp^\infty$ (surrounding the disk sections of the strings of finite area \eqref{A}).
    The horizon may be considered as the starting surface of outgoing  lines of force which  propagate partly to infinity, and partly to Misner strings, where they turn back to the horizon, forming a non-uniform singular flux. They then spread along the horizon and reemerge outward. These closed lines balance the difference of the GE flow through the infinite sphere, showing the presence of the total mass $m$ inside, and the outgoing flux from the horizon, showing the total mass $\sqrt{m^2+n^2}$. The excess on $\sqrt{m^2+n^2}-m$ corresponds exactly to the flow through the lateral surface $S^\pm_L$, so this part of flux  is circulating. We depict surfaces at which the flux is outgoing by red color symbolizing the positive charge, and by blue, when the flux is ingoing. Recall, that Misner strings are two-dimensional objects with non-zero transverse size \eqref{A}. The line going to infinity can be considered as closing there to reemerge on the entering sections $S_\perp^\infty$ of MS through which they are transported back to the horizon. } 
    \label{fig:GE}
 \end{figure}

The mass contained inside the sphere $S_r$ of the radius $r$ is given
by
\begin{equation}\label{Mr}
M_r=\frac1{8\pi}\Phi_r=m\frac{\tilde{\Delta}}{\Sigma},
\end{equation}
so the total mass $M_\infty=m$ is less then the horizon mass $M_H=\sqrt{m^2+n^2}$, the difference being gives by the flux {\em ingoing} through the lateral surfaces of north and south Misner strings:
\begin{equation}
M_\infty-M_H=m-\sqrt{m^2+n^2}=\frac1{8\pi}\left(\Phi_L^++\Phi_L^-\right). 
\end{equation}
The picture of GE lines of force in Taub-NUT solution schematically is presented in Fig.\ref{fig:GE} and will be illustrated in a number of exact plots in Sec.\ref{sec: hair patterns}.

\begin{figure}[H]
\centering\includegraphics[width=0.79\linewidth]{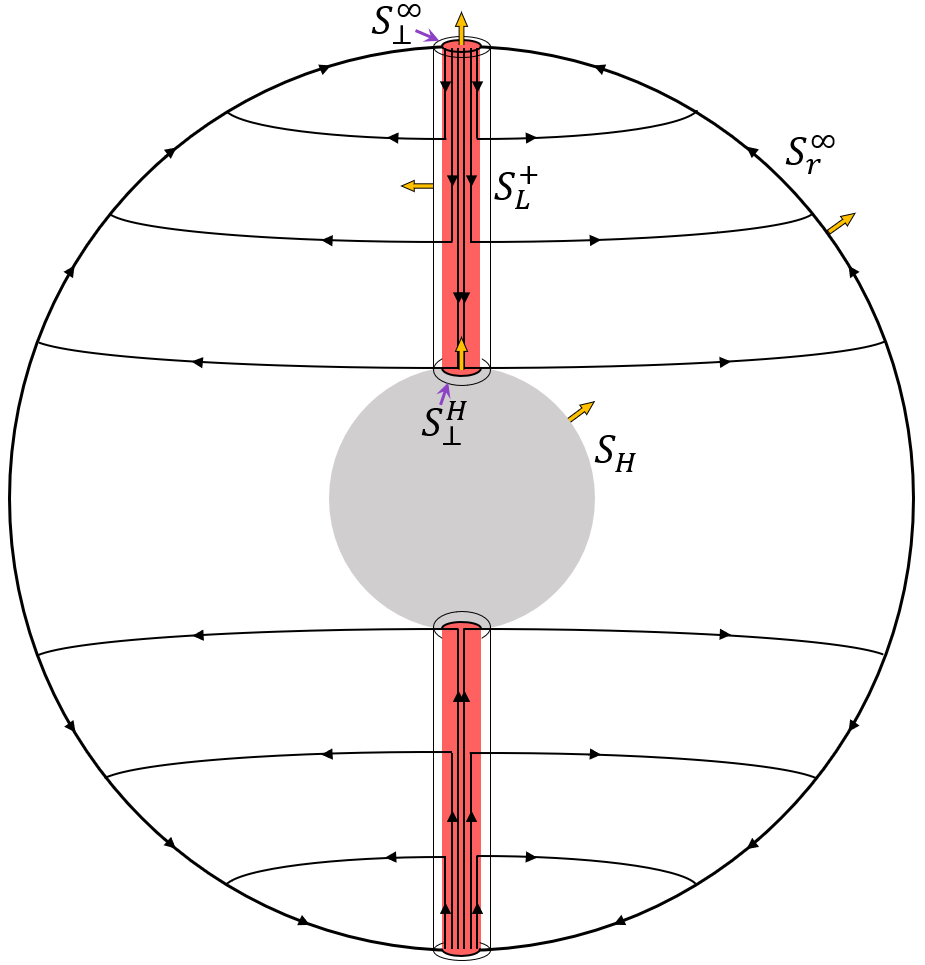}
\vspace{-0.3 cm}
 	\caption{Schematic picture of GM lines of force. These are entering through Misner strings at infinity and then gradually turn to the bulk through the lateral surfaces, fully expiring at the horizon. Misner strings are marked red since they serve the source of GM lines in the bulk which propagate to infinity, closing there at the entrance circles of the strings.}
	\label{fig:GM}
 \end{figure}

One can perform similar calculations for the gravimagnetic fluxes, obtaining: 
 \begin{align}
&\tilde{\Phi}_\perp^\pm(\infty) =-4\pi n,\\
&\tilde{\Phi}_\perp^\pm(r_H)=0,\\
&\tilde{\Phi}_L^\pm =4\pi n.
 \end{align}
 One can see that Misner strings have zero gravimagnetic mass as well.
 The GM flux in the bulk though the sphere $S_r$ is equal to
 \begin{equation}\label{Nr}
N_r=\frac1{8\pi}\tilde{\Phi}_r=n\frac{{\Delta}}{\Sigma},
\end{equation}
so the GM flux from the horizon is equal to zero while the total flux through the infinite sphere $N_\infty =n$. In this case the flux at infinity is entirely due to the outgoing fluxes through the lateral surfaces of Misner strings. This is shown schematically in Fig.\ref{fig:GM} and will be presented in exact plots later on.

In short, the Taub-NUT metric is a solution without sources of physical mass or gravimagnetic mass, which can be understood as an engine transporting GE and GM fluxes from infinity to the horizon and back. The field lines are closed, although not smooth: they rotate by $\pi/2$ when entering or escaping to the  bulk. The propagation of GE field lines  entering the horizon out of MS is smooth, a situation analogous to the propagation of Coulomb field lines from a charge approaching the horizon \cite{Hanni:1973fn}.

\newpage
\vspace{-0.5 cm}
\vspace{-0.8 cm}

The correspondence of the quantities introduced here with similar ones considered by us in electrodynamics \cite{Galtsov:2026ndi} is presented in Table \ref{tab: em/gr}.
 \begin{table}[H]
 \centering
 	\begin{tabular}{|c|c|}
 		\hline 
 		&\\
 		 \hspace{0.3 cm}ELECTRODYNAMICS\hspace{0.3 cm}  &  \hspace{0.3cm}GRAVITY\hspace{0.3cm} \\[7pt]
 		\hline 
 		Potential: & Komar one-form:\\[6pt]
 		
 		$A=A_\mu dx^\mu$ & $k=g_{t\mu}dx^\mu$\\[6pt]
 		\hline
 		Field strength: &Komar two-form:\\[5pt]
 		$F=dA$ & $K=dk$\\[6pt]
 		\hline
 		Bianchi identities: & Bianchi identities:\\[5pt]
 		$dF=0$ & $dK=0$\\[6pt]
 		\hline
 		Singular magnetic flux: & Singular GM flux:\\[6pt]
 		$\mathcal{F}=2\pi A_\varphi^\pm \delta^2(\mathbf{x})dx\wedge dy$ & $\mathcal{K}=2\pi g_{t\varphi}^\pm \delta^2(\mathbf{x})dx\wedge dy$\\[5pt]
 		\hline
 	\end{tabular}
    \caption{Similar quantities in electrodynamics and gravity. }
    \label{tab: em/gr}
 \end{table}

 \section{Kinks and fictitious charges}
 
The picture presented above considers the Taub-NUT solution to be a true vacuum solution. This corresponds to the absence of a central singularity where potential sources of mass and gravimagnetic mass could be located, and presents the metric as describing a peculiar gravitational soliton with a horizon. However, this picture comes at the cost of the nonsmooth nature of the GE and GM field lines. The bulk field lines are orthogonal to the string, while the singular field is collinear with the strings, and these two facts are inextricably linked to the underlying geometry. Meanwhile, not only the integral fluxes of the bulk fields through the lateral surfaces are balanced with the fluxes through the string, but, as we will see shortly, the balance is maintained locally at any distance from the horizon along the strings. So the field lines, although continuous, inevitably have kinks with the angle $\pi/2$ when entering or leaving the strings. These kinks produce non-zero divergence of the internal lines. Indeed, the fluxes \eqref{NF} and \eqref{MF} are non-uniform and depend on the coordinate $z$ along the strings. So we have some three-dimensional vector depending on the coordinate along it, like $E_z(z)$ say, which has non-zero divergence $div{\bf E}=dE_z/dz$. These quantities can be regarded as densities of the fictitious charges.

Performing the calculations, one obtains for fictitious densities of mass and NUT charges along the strings the following expressions:
\begin{equation}\label{TaubNUT mass and NUT distr}
    	\lambda_m(r)=-n^2\frac{\Delta}{\Sigma^2},\quad
\lambda_n(r)=mn\frac{\tilde{\Delta}}{\Sigma^2}.
    \end{equation}
     These   tend to zero at infinity. The Misner string masses calculated using Komar's lateral integrals in \cite{Manko:2005nm,Clement:2017otx,Clement:2019ghi} can be rediscovered by integrating $\lambda_m(r)$ over the strings:
     \begin{equation}
         M^\pm_{\rm fict}=\frac{1}{8\pi}\Phi_L^\pm=\int_{r_H}^\infty \!\!\!\lambda_m dr=\frac12(m-\sqrt{m^2+n^2}).
     \end{equation}
    
Similarly, fictitious mass densities $\lambda_m^H$ and NUT densities $\lambda_n^H$ of the horizon can be introduced using the definitions
\begin{equation}\label{def  HorizonDensities}
  M_H=\int_H \lambda^H_m\sqrt{\gamma_H}  d\theta d\varphi, \;\; N_H=\int_H \lambda^H_n \sqrt{\gamma_H} d\theta d\varphi. 
\end{equation}
This leads to the expressions
\begin{align}
		&\lambda_m^H=\frac{1}{4\pi(r_H^2+n^2)}\left(m+2n^2\frac{\sqrt{m^2+n^2}}{r_H^2+n^2}\right),\;\;\;\lambda_n^H=0,
	\end{align}
   where the NUT density is identically zero, while the mass density remains constant all over the horizon. This is true only for non-rotating solutions.

\section{S-duality}

One can check that discrete S-duality
\begin{equation}\label{grav discr S dual}
	m\rightarrow n,\qquad n \rightarrow -m,
    \end{equation}
hold for the Komar two-form and its dual in the limit $r\to\infty$
\begin{equation}
    	K\rightarrow2n\sin \theta d\theta \wedge d\varphi,\quad
    	\star K \rightarrow -2m\sin \theta d\theta \wedge d\varphi.
    \end{equation}
This corresponds to linearized duality.

Linearizing the Taub-NUT metric $g_{\mu\nu}\approx\eta_{\mu\nu}+h_{\mu\nu}$ and expressing the linear part $h_{\mu\nu}$ in  Cartesian coordinates one obtains the relation
    \begin{align}
    	h_{tt}=-\frac{2m}{r},\quad h_{ti}=\frac{2nz\epsilon_{zij}x^j}{r(r^2-z^2)}\quad h_{ij}=\frac{2mx_ix_j}{r^3},
    \end{align}
    where $r=\sqrt{x^ix_i}$. Then the linearized one-form $k$ is expressed as
    \begin{align}
    k=-fdt-\frac{2nz\epsilon_{zij}x^idx^j}{r(r^2-z^2)},
    \end{align}
where $f=1-2m/r$.
 Calculating exterior derivative one obtains for the Komar two-form  the relation
    \begin{align}
    	K=\frac{2m}{r^3}x_idt
    \wedge dx^i+\frac{n}{r^3}\epsilon_{ijk} x^idx^j\wedge dx^k,
    \end{align}
    taking the Hodge dual one sees that the quantity $\star K$ can be obtained from $K$ replacing mass and NUT in accordance with \eqref{grav discr S dual},
 confirming that the field tensors $K_{\mu\nu}$ and $\tilde{K}_{\mu\nu}$  become dual to each other in linearized level. 

    Meanwhile, the fluxes of GE and GM field lines through the sphere $S_r$ in the bulk are exactly S-dual. Indeed, comparing the formulas \eqref{Mr} and \eqref{Nr} one finds
 \begin{equation}
   M_r^2+N_r^2=m^2+n^2 
 \end{equation} which is invariant under general $SO(2)$ duality rotation
 \begin{equation}\label{grav s duality}
	\begin{aligned}
		m&\rightarrow m \cos \alpha + n \sin \alpha\\
		n&\rightarrow n \cos \alpha-n\sin \alpha.
	\end{aligned}
\end{equation}
 The aspects of   gravitational duality in the full nonlinear theory and its realization on the linearized level were extensively discussed in the literature \cite{Ellwanger:2001uq,Deser:2005sz, Argurio:2008zt,Argurio:2009xr,Kol:2020zth,Kol:2022bsd,Kol:2023yxd,Hull:2000zn,Hull:2001iu,Hull:2023iny,Moutsopoulos:2009ia,Dehouck:2011xt}.

\section{Rotation}
Rotating generalization is presented by the Demianski-Newman solution  \cite{Demianski66}:
		\begin{equation*}\label{KN metric}
			ds^2=-\frac{\Delta}{\Sigma}(dt-\beta d\varphi)^2+\frac{\sin^2\theta}{\Sigma}(\alpha d\varphi-adt)^2+\frac{\Sigma}{\Delta}dr^2+\Sigma d\theta^2,
		\end{equation*}
        where
        \begin{align*}
            &\Delta=r^2-2mr+a^2-n^2,\;\; \Sigma=r^2+\nu^2,\;\; \nu=n+a\cos \theta,\\
				&\alpha=r^2+n^2+a^2,\quad\beta=a \sin^2 \theta-2n\cos \theta.
        \end{align*}
		Following the same lines as for the static case one obtains 
the Komar one-form
\begin{equation}\label{KerrNUT 1-form}
    k=-\frac{\Delta-a^2\sin^2\theta}{\Sigma}dt+\frac{\beta \Delta-a \alpha \sin^2 \theta}{\Sigma}d\varphi.
\end{equation} The Komar two-form  $K=dk$ in the bulk is rather cumbersome and will not be given. We present just the potential $ \tilde{k}$ generating dual two-form
\begin{equation}\label{KerrNUT dual 1-form}
 \tilde{k}=\frac{m}{n}\left[\frac{\tilde{\Delta}-a^2\sin^2\theta}{\Sigma}dt-\frac{\beta \tilde{\Delta}-a \alpha \sin^2 \theta}{\Sigma}d\varphi\right],
\end{equation} where we now have
\begin{equation*}
\tilde{\Delta}=r^2+2r\frac{n^2}{m}+a^2-n^2.
\end{equation*}
Together with \eqref{KerrNUT 1-form} this is needed to calculate singular fluxes in Misner strings.

Differentiating the one-forms \eqref{KerrNUT 1-form}, \eqref{KerrNUT dual 1-form} in terms of distributions, one obtains the singular fluxes in the Misner strings
\begin{align}
   	& \tilde{\mathcal{K}}^\pm= \pm m \frac{ 4\pi\tilde{\Delta}\delta(x)\delta(y)}{r^2+(n\pm a)^2}dx \wedge dy,\\
   	&\mathcal{K}^\pm= \mp n\frac{4\pi \Delta\delta(x)\delta(y)}{r^2+(n\pm a)^2}dx \wedge dy.
\end{align}
Their novel features are that now the magnitudes of these fields are different for upper and lower strings, due to  rotation.
As in the static case, the gravimagnetic fluxes vanish on the horizon.
 
Let's check the balance equation for the GE field for the north string verifying that its mass is zero. The fluxes are
\begin{align}
	&\Phi_\perp^+(\infty) =-4\pi m,\\
	&\Phi_\perp^+(r_H)=-4\pi\frac{am+n\sqrt{m^2+n^2-a^2}}{n+a},\\
	&\Phi_L^+ =4\pi n\frac{m-\sqrt{m^2+n^2-a^2}}{n+a},
\end{align}
and the balance equation for the upper string
\begin{equation}
	\Phi_\perp^+(\infty)-\Phi_\perp^+(r_H)+\Phi_L^+=0,
\end{equation}
is again satisfied. It can also be shown that such relation holds also for the lower string, so the statement that the Misner strings are massless is also true in the case of rotation. One can also check the analogous equations for the bulk fluxes and for the GM field.

 One can also introduce the fictitious densities of the mass and NUT charge on the strings and the horizon as in the static case. The calculation of the linear densities on the strings give
\begin{align}
    &\lambda_m^\pm=-n\frac{nr^2-2mr\nu_\pm-n\nu_\pm^2}{(r^2+\nu_\pm^2)^2},\\
	&\lambda_n^\pm=n\frac{mr^2+2nr\nu_\pm-m\nu_\pm^2}{(r^2+\nu_\pm^2)^2},
\end{align}
where  $\nu_\pm=n\pm a$. Note that now the densities of mass and NUT become different for north and south strings, so the configuration becomes asymmetric with respect to the equatorial plane. 
The horizon densities defined by \eqref{def  HorizonDensities} are now given by the expressions
\begin{align}
&\lambda_m^H=\frac{m(r_H^2-\nu^2)+2n\nu r_H}{4\pi(r_H^2+\nu^2)^2},\\
	&\lambda_n^H=-a\frac{2(mr_H+n^2)\cos \theta+an(1+\cos^2\theta)}{4\pi(r_H^2+\nu^2)^2}.
\end{align}

If one considers a pure Kerr solution, i.e. the case $n=0$, then the Komar two-form can be presented as
\begin{equation}
	\begin{aligned}
		K=-\frac{2m}{\rho^2}\big[&(r^2-a^2\cos^2 \theta) dr\wedge(dt-a\sin^2\theta d\varphi)-\\
		&-ar\sin2\theta\; d\theta\wedge(adt-(r^2+a^2)d\varphi)  \big],
	\end{aligned}
\end{equation}
where $\rho=r^2+a^2\cos^2\theta$. The component $K_{\theta \varphi}$, responsible for the flux of GM field through the sphere changes sign as one crosses equatorial plane. Thus, considering the sphere corresponding to the surface of the horizon we obtain, that the force lines change their direction, so the horizon's surface should be polarized. Also, the NUT density on the horizon in the limit $n\to0$ reduces to
\begin{equation}
    \lambda^H_n=-\frac{amr_H\cos \theta}{2\pi(r_H^2+a^2\cos^2\theta)^2},
\end{equation}
so the NUT density changes its sign on the equatorial plane.

The GM polarization of the horizon is clearly seen in  the force line plots for the Kerr solution in Sec.\ref{sec: hair patterns}.

\section{Lines of force}
 
\subsection{Definition of the lines of force}

Following the idea proposed by Christodoulou and Ruffini for the electric and magnetic fields \cite{Christodoulou:1973}(see also \cite{Mcguire:1975wq}, \cite{Hanni:1973fn, Mcguire:1975wr,Pizzi:2007zz, Paolino:2008qi}) we analogously define a GE and GM force lines(FLs) as lines, tangent to these fields as measured by a  ZAMO-observer with 4-velocity:
\begin{equation}\label{ZAMO 0}
	u_\mu=(-V,0,0,0),
\end{equation}
where $V=\sqrt{-1/g^{tt}}$ is the lapse function.
Then the components of these fields will read
\begin{equation}
	G^\mu=\frac{1}{2}u_tK^{t\mu},\qquad
	\tilde{G}^\mu=\frac{1}{2}u_t\tilde{K}^{t\mu}.
\end{equation} 
We then introduce the GE force lines(analogously for GM) as the lines tangent to these fields $dr/d\lambda=G^r$ and $d\theta/d\lambda=G^\theta$,
with $\lambda$ being the parameter along a FL. Such definition is equivalent to the FLs, defined as the locus of points with fixed value of the flux $\Phi$ \cite{Hanni:1973fn}. The slope of the lines in this definition is given by
\begin{equation}
	\frac{dr}{d\theta}=-\frac{\partial \Phi/\partial \theta}{\partial \Phi/\partial r}=\frac{\sqrt{-g}K^{tr}}{\sqrt{-g}K^{t\theta}},
\end{equation} 
which indeed defines the same form of curves.

We also introduce \textit{compactified} Cartesian-like coordinates $x^1$ and $x^2$ by replacing $r$ with $\arctan(r/k)$ and considering $\theta$ as the polar angle. Here $k$ is some constant that we use to adjust the size of the horizon on the pictures. Such transformation brings spatial infinity $r\rightarrow\infty$ to a finite circle of radius $\pi/2$.
 
On the pictures of FLs, presented below, we have a following conventions. The red(blue) color represents the positions of the positive(negative) fictitious masses(or NUTs). The horizon is drawn by the gray disk in the center and the circle around it represents the sign of the induced fictitious mass(NUT). The black large circle corresponds to the infinitely large sphere. 

\subsection{Physical interpretation of gravielectric field}

Previously, we introduced the GE field, taking advantage of the similarity of the formulas and equations with electrodynamic ones \cite{Galtsov:2026ndi}. In this subsection we will show that in fact the formula we introduced has a direct physical meaning associated with the characteristics of the gravitational field. The 4-velocity of the ZAMO-observer \eqref{ZAMO 0} is
\begin{equation}
	u=\frac{1}{V}(\partial_t+\Omega \partial_\varphi),
\end{equation}
where $\Omega=g^{t\varphi}/g^{tt}$ being the angular velocity relative to distant stars. Then for the  4-acceleration one obtains  
\begin{equation}
	a^\mu=\frac{1}{V}(G^\mu+\Omega R^\mu),
\end{equation}
where we have introduced the quantity $R^\mu=u_t\nabla^t\delta^\mu_\varphi$, associated with the rotational Killing vector $\partial_\varphi$.
The first term is identified with the gravitational force, and the second is an analog of the Coriolis force.
The acceleration experienced at infinity is then given by
\begin{equation}
	a_\infty=V\sqrt{a_\mu a^\mu}=|G^\mu+\Omega R^\mu|.
\end{equation}
If one now considers the static spacetime,  the concept of a static observer becomes clearly defined as $\Omega\equiv0$, and the acceleration evaluated on the horizon coincides exactly with the \textit{surface gravity}:
\begin{equation}
	G_H\equiv\sqrt{G_\mu G^\mu}\Big|_{r=r_H}=a_\infty(r_H)=\kappa,
\end{equation}
thus confirming the physical significance of the GE field $K^{\mu\nu}$ as the \textit{gravitational field strength}.

\subsection{GE and GM multipole moments}
In the paper \cite{Mukherjee:2020how} it was shown that using the Geroch-Hansen approach to obtain the gravitational multipole moments for the Kerr-NUT solution, one can find that they are given by the formula
\begin{equation}
    \mathbb{M}_n=(m-in)(ia)^n,
\end{equation}
so for the mass part   one has
\begin{equation}\label{mass multipole}
    M_0=m,\quad M_1=na,\quad M_2=-ma^2,\quad  \dots \;\;.
\end{equation}
Using now the fact that these moments are obtained by taking the limit $r\rightarrow \infty$ and the fact that the gravitational duality is preserved in the linear level, one can introduce dual gravitational multipole moments as
\begin{equation}
    \mathbb{N}_n=(n+im)(ia)^n,
\end{equation}
so their NUT parts at this time will be given by
\begin{equation}\label{nut multipole}
    N_0=n,\quad N_1=-ma,\quad N_2=-na^2,\quad  \dots \;\;.
\end{equation}
From these expressions, we can obtain several rather interesting special cases. For
the case of pure mass $a=n=0$, i.e. the Schwarzschild solution, from  \eqref{mass multipole} we see that only the monopole parts are present, while all moments associated with the NUT are absent.
If we now add rotation while leaving the NUT parameter equal to zero, then quadrupole components appear in the mass moments, while the dipole component remains equal to zero. Thus, for the mass, the field lines should also extend to infinity, and be curved due to rotation. For the NUT moments, terms appear, starting with the dipole component, and so on. Thus, we expect the NUT field lines to be all closed.
Next, we can consider the case of a pure NUT. Then, for the mass, all moments are absent, and the NUT is left with a monopoly part.

Finally, in the general case, where both mass with NUT and rotation are present, both fields possess all multipole moments, and this results in an asymmetry relative to the equatorial plane as was also noted in \cite{Mukherjee:2020how}.

\newpage
\subsection{GE and GM field patterns}\label{sec: hair patterns}

\begin{figure}[H]
 	\centering
\includegraphics[width=0.99\linewidth]{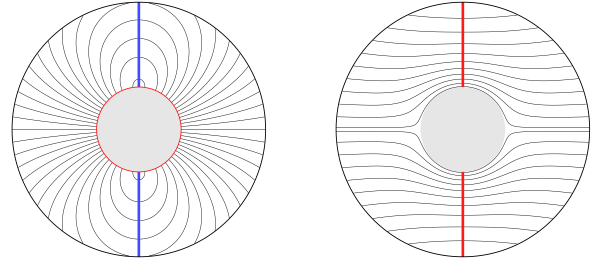}
 	\caption{{\bf$m=1,\; n=1,\; a=0.$} Taub-NUT solution. The figure for gravielectric line (left panel) shows how the excess mass of the horizon $\sqrt{m^2+n^2}-m$ is returned back through the outgoing flux from the horizon   to Misner strings, which are transparent in the transverse direction. The remaining lines tend to infinity, carrying a flux proportional to $m$. Gravimagnetic flux (right panel) comes from infinity through the Misner strings, and propagates though their lateral surfaces to infinity, where it closes again on the entering circles of Misner strings. GM flux vanishes at the horizon}
 	\label{m=1 n=1 a=0}
 \end{figure}

\begin{figure}[H]
	\centering
\includegraphics[width=0.99\linewidth]{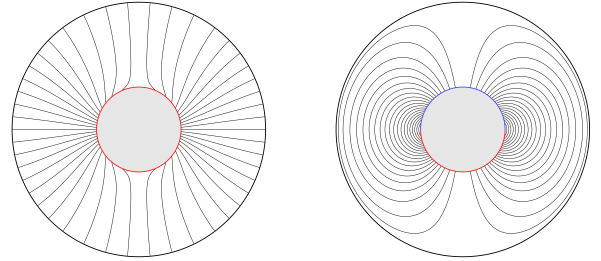}
 	\caption{{\bf$m=1,\; n=0,\; a=1.$} This is Kerr solution without Misner strings. GE lines (left panel) propagate from the horizon (which is positively charged) to infinity. GM lines (right panel) form  dipole picture (which is exact, not in dipole approximation). The upper half of the horizon is negatively charged with respect to GM mass, the lower is positively charged, the separation being induced by rotation. GM lines start from positively charged horizon hemisphere and end on the negatively charged one. }	\label{m=1 n=0 a=0.5}
\end{figure}

\begin{figure}[H]
 	\centering
\includegraphics[width=0.99\linewidth]{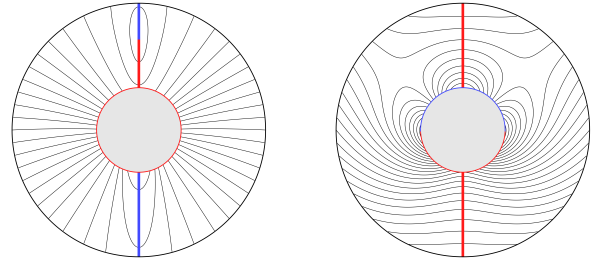}
 	\caption{{\bf$m=1,\; n=0.2,\; a=0.5.$} NUT and rotation of the same order. The horizon is GE positive, emitting the GE lines, together with part of the north MS. The south MS is negatively charged (absorbing). There are confined (non propagating to infinity) field lines between the horizon and the south string and string-string line on the north MS. The lines escaping to infinity emerge on the horizon and the positively charged part of the north MS. GM lines form confined structures of two kinds: horizon-horizon (dipole-like) and north string-horizon. The remaining lines reaching infinity start either on the horizon or from both MS, which are positively GM charged.}
 	\label{m=1 n=0.2 a=0.5}
 \end{figure}

\begin{figure}[H]
  	\centering
\includegraphics[width=0.99\linewidth]{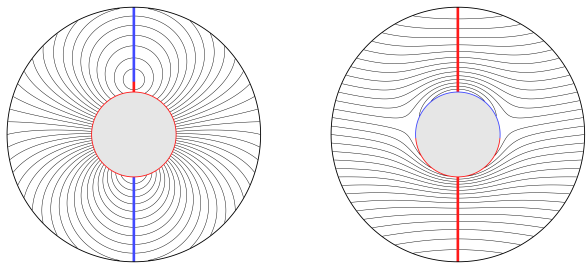}
  	\caption{{\bf$m=1,\; n=1.5,\; a=0.5.$} Taub-NUT with a large NUT-parameter. GE picture (left) consist of lines going to infinity from the horizon and confined lines of horizon-string and string-string type. GM pattern (right) contains string-infinity and horizon-infinity lines and confined sting-horizon lines. }
  	\label{m=1 n=1.5 a=0.5}
 \end{figure}

\begin{figure}[H]
  	\centering
\includegraphics[width=0.99\linewidth]{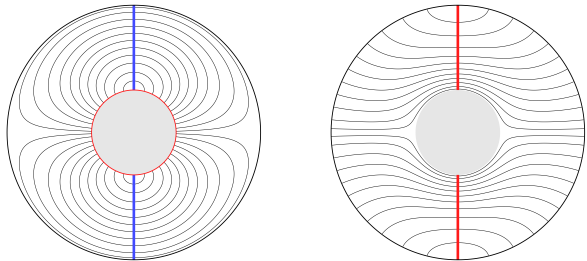}
 	\caption{{\bf$m=0,\; n=1,\; a=0.$} Pure  NUT  non-rotating solution. No GE lines going to infinity, but confined lines starting from the horizon, penetrating to both MS via lateral surfaces and returned back to the horizon by singular fluxes. GM lines (right panel) initiate on MS and propagate to infinity, where they close at the entering circles of MS. These are circulating between infinity and MS, not touching the horizon. }
 	\label{m=0 n=1 a=0}
\end{figure}

\begin{figure}[H]
 	\centering
\includegraphics[width=0.99\linewidth]{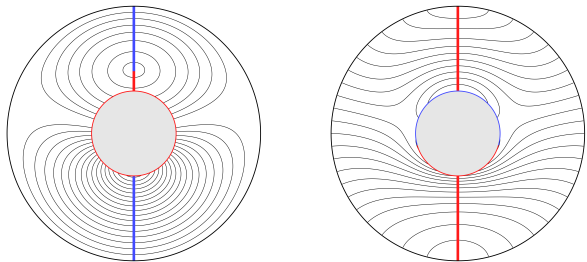}
 	\caption{{\bf$m=0,\;n=1,\; a=0.5.$} Pure NUT-solution with rotation. No GE lines (left panel) reaching infinity; confined lines of horizon-string and string-string type. GM lines (right panel) partly reach infinity and partly confined (string-horizon type)}
 	\label{m=0 n=1 a=0.5}
 \end{figure}

 Here we present the GE and GM field line patterns Figs.\ref{m=1 n=1 a=0}-\ref{m=0 n=1 a=0.5}  for different values of mass, NUT and rotation. 
 \vspace{-0.3 cm}
\vspace{-0.6 cm}
\vspace{-0.6 cm}
 \vspace{-0.5 cm}

   \section{Conclusions} 
   
   We revisited structure of Misner strings in solutions with magnetic mass emphasizing that these are two-dimensional defects, which are better visualized not as rotating rods, but as tubes with  finite lateral area per unit length. Introducing gravielectric and gravimagnetic fields with the help of the Komar two-form and its Hodge-dual  we found that Misner strings  of Taub-NUT and Kerr-Taub-NUT solutions  carry singular GE and GM fluxes. These fluxes, together with bulk field lines, form closed contours of two kinds: confined (horizon-string and string-string type) and non-confined which propagate to infinity, where they are closing to reemerge at the Misner strings opening circles and returned back to the horizon.
   The boundary of the Misner string consist not only of the lateral surface but also includes transverse sections at infinity and at the horizon through which the singular GE and GM fields have finite fluxes. Summing up fluxes through all boundaries of Misner strings, we prove that these are massless, as Bonnor assumed in his original proposal. We present typical patterns of GE and GM fields of nutty black holes with different sets of parameters.


   \section{Acknowledgments}
  The authors thank G\'erard Cl\'ement for useful comments and discussions.
 The work  was supported by the Foundation for the Advancement of Theoretical
Physics and Mathematics 'BASIS'.

\end{document}